\documentclass{aa}  

\usepackage{graphicx}
\usepackage{txfonts}
\usepackage{multirow}
\usepackage{CJKutf8}

\defcitealias{2016ApJ...817...34M}{M16}
\newcommand\micron{\mbox{$\mu$m}}

\begin{document}

   \title{Identifying Compton-thick AGNs in the COSMOS}

   \subtitle{I. Among X-ray AGNs with low photon counts}

   \author{Xiaotong Guo \begin{CJK}{UTF8}{gkai}(郭晓通)\end{CJK}
          \inst{1,2}
          \and
          Qiusheng Gu \begin{CJK*}{UTF8}{gkai}(顾秋生)\end{CJK*}\inst{3,4}\fnmsep$^\star$
          \and 
          Guanwen Fang \begin{CJK*}{UTF8}{gkai}(方官文)\end{CJK*}\inst{1,2}
          \and
          Yongyun Chen \begin{CJK*}{UTF8}{gkai}(陈永云)\end{CJK*}\inst{5}
          \and
          Nan Ding \begin{CJK*}{UTF8}{gkai}(丁楠)\end{CJK*}\inst{6}
          \and
          Xiaoling Yu \begin{CJK*}{UTF8}{gkai}(俞效龄)\end{CJK*}\inst{5}
          \and
          Hongtao Wang \begin{CJK*}{UTF8}{gkai}(王洪涛)\end{CJK*}\inst{7}
          }

   \institute{School of Mathematics and Physics, Anqing Normal University, Anqing 246133, China
                 \and
                         Institute of Astronomy and Astrophysics, Anqing Normal University, Anqing 246133, China\\
                         \email{guoxiaotong@aqnu.edu.cn}
         \and
             School of Astronomy and Space Science, Nanjing University, Nanjing, Jiangsu 210093, China\\
             \email{qsgu@nju.edu.cn}
         \and
             Key Laboratory of Modern Astronomy and Astrophysics (Nanjing University), Ministry of Education, Nanjing 210093, China
         \and
                College of Physics and Electronic Engineering, Qujing Normal University, Qujing 655011, China
         \and
                School of Physical Science and Technology, Kunming University, Kunming 650214, China
         \and
                School of Science, Langfang Normal University,Langfang 065000, China
             }

   \date{Received 30 August 2024; accepted 15 January 2025}

 
  \abstract
   {Compton-thick active galactic nuclei (CT-AGNs), characterized by a significant absorption with column densities of $\mathrm{N_H}\geqslant 1.5\times 10^{24} \ \mathrm{cm}^{-2}$, emit feeble X-ray radiation and are even undetectable by X-ray instruments, making them difficult to identify. X-ray radiation from AGNs is the predominant source of the cosmic X-ray background (CXB). Based on AGN synthesis models for the CXB, the fraction of CT-AGNs should constitute a substantial portion of AGN population, 
   	approximately 30\% or more.}
   {The fraction of CT-AGNs discovered in the Cosmological Evolution Survey (COSMOS) is significantly lower than this value. This means that many CT-AGNs may be hidden in AGNs that exhibit low photon counts or that have not been detected by X-ray instruments. This work focuses on identifying CT-AGNs hidden in AGNs with low photon counts.}
   {Firstly, we selected 440 AGNs with abundant multiwavelength data as our sample. Secondly, we analyzed multiwavelength data, extracting crucial physical parameters required for the CT-AGN diagnosis. Finally, we used multiwavelength approaches to identify CT-AGNs.}
   {We have successfully identified 18 CT-AGNs in our sample. Among the CT-AGNs, four AGNs show discrepant results across different diagnostic methods. We discuss the potential reasons behind these diagnostic discrepancies. We explore the impact of estimating [O~III]$\lambda~5007$ luminosities based on [O~II]$\lambda~3727$ luminosities for the CT-AGN diagnosis. We have also found that the properties of host galaxies for CT-AGNs and non-CT-AGNs do not show significant discrepancies.}
   {}
   

   \maketitle
%

\section{Introduction}

   Active galactic nuclei (AGNs) are some of the most luminous and persistent sources in the Universe, with their emission covering almost the entire electromagnetic spectrum. The power of AGNs originates from the accretion of surrounding matter by the central supermassive black holes. The emissions across various wavelengths arise from the distinct structures of AGNs. 
   For example, the emission from ultraviolet (UV) to optical is mostly thermal radiation, originating from the accretion disk \citep[e.g.,][]{1980ApJ...235..361R}; the mid-infrared (MIR) emission is mainly the reradiation of the torus after absorbing UV and optical radiation \citep[e.g.,][]{1993ARA&A..31..473A}; and the X-ray emission predominantly arises from multiple inverse Compton up-scatterings of accretion disk photons by hot electrons in the corona \citep[e.g.,][]{1979ApJ...234.1105L,1991ApJ...380L..51H}. The emissions from the central engine (including the accretion disk and corona) of the AGNs can be absorbed by the surrounding torus, which is composed of dense dust and gas \citep[e.g.,][]{1993ARA&A..31..473A,1995PASP..107..803U,2015ARA&A..53..365N}. As a result, depending on different viewing angles for the torus, part of the emission from the central engine can be suppressed with varying intensities. 
   When our viewing angle is “edge-on” (i.e., the emission from the central engine passes through the torus), the UV to optical emission will be extinguished. Although the X-ray emission has a strong ability to pierce through, part of it is still absorbed.
   The AGNs with column densities ($\mathrm{N_H}$) of the absorbed gas exceeding $1.5\times 10^{24}\ \mathrm{cm}^{-2}$, a high level of absorption, are classified as Compton-thick AGNs \citep[CT-AGNs; e.g.,][]{2004ASSL..308..245C}.
   
   The luminous and persistent X-ray emission is an important observational feature that distinguishes AGNs from other extragalactic sources.
   Moreover, the X-ray radiation from AGNs is recognized as the predominant contributor to the cosmic X-ray background (CXB), with up to 93\% of the extragalactic CXB at energies below 10 keV having been resolved into AGNs \citep[e.g.,][]{2003ApJ...588..696M, 2005MNRAS.357.1281W, 2023ExA....56..141L}. 
   CT-AGNs, a subclass of AGNs, are also significant contributors to the CXB. In particular, a considerable fraction of emission (10\%--25\%) is expected to be produced by a large population of CT-AGNs at the peak of the CXB \citep[$\sim$ 30~keV;][]{2008ApJ...689..666A}. 
   \cite{2011ApJ...728...58B}  modeled and analyzed the CXB using X-ray radiation from AGNs with varying absorption levels, recovering the intrinsic distribution of absorbing column densities, and demonstrated that CT-AGNs are $20^{+9}_{-6}\%$ in the intrinsic AGN population.
   In some models, a higher fraction of CT-AGNs is required, even reaching 50\% of the total AGN population \citep[e.g.,][]{2014ApJ...786..104U, 2015ApJ...802...89B, 2019ApJ...871..240A}.
   However, the observed fraction of CT-AGNs within the total AGN population in the local Universe is 5\%--10\% \citep[z$\leqslant$0.1; e.g.,][]{2015ApJ...815L..13R}. 
   The observed fraction of CT-AGNs is so low that their X-ray emissions cannot adequately explain the peak of the CXB. This also suggests that a large number of CT-AGNs remain unidentified or undiscovered.
   
   To identify the missing CT-AGNs in the local Universe, the X-ray astronomy group at Clemson University and the extragalactic astronomy group at the INAF-OAS Bologna and INAF-IASF Palermo initiated a targeted project in 2017 to actively search for CT-AGNs, known as the Clemson-INAF CT-AGN Project \citep[CCTAGN;][]{2018ApJ...854...49M}. The CCTAGN project is built upon a volume-limited sample of AGNs detected by BAT during 150 months of observations within the redshift range of $z<0.05$. This sample is almost unbiased and is used to determine the actual fraction of CT-AGNs within the AGN population. 
   This project has made a series of impressive research findings \citep[e.g.,][]{2018ApJ...854...49M, 2019ApJ...871..182Z, 2019ApJ...870...60Z, 2019ApJ...882..162M, 2019ApJ...872....8M, 2021ApJ...922..159T, 2021ApJ...922..252T,2022ApJ...935..114M,2022ApJ...940..148S}, with a total of 35 CT-AGNs discovered, constituting approximately 8\% of the AGN sample \citep{2023A&A...676A.103S}. 
   The fraction of CT-AGNs recovered within $z<0.01$ is $20\% \pm 5\%$ \citep{2021ApJ...922..252T}, which aligns with the predicted fraction from the AGN synthetic model of the CXB. This suggests that the CXB in the local Universe within $z<0.01$ is well explained.
   
   Due to the dense gas environments in high-redshift galaxies \citep[e.g.,][]{2013ARA&A..51..105C}, the emission from AGNs at their centers is more likely to be significantly absorbed. Hence, we anticipate that the fraction of CT-AGNs in high redshift environments is higher than that in the local Universe. In fact, the fraction of identified CT-AGNs consistently falls short of the theoretical expectations derived from CXB models in numerous X-ray deep field surveys. For instance, in the Chandra Deep Field-South survey, 711 AGNs have been identified by X-ray \citep{2017ApJS..228....2L}, of which 79 are certified as CT-AGNs \citep[e.g.,][]{2017ApJS..232....8L, 2019A&A...629A.133C, 2019ApJ...877....5L, 2020ApJ...903...49L, 2021ApJ...908..169G}, accounting for 11\%.
   Even the Nuclear Spectroscopic Telescope Array \citep[NuSTAR,][]{2013ApJ...770..103H}, which is sensitive above 10 keV, \cite{2018ApJS..235...17M} has only been able to increase the fraction of CT-AGNs to approximately 11.5\% in the UKIDSS Ultra Deep Survey field.
   
   To further enhance the fraction of CT-AGNs at high redshifts and approach the theoretical predictions of the CXB, it is necessary to search for CT-AGNs in diverse X-ray deep field surveys. 
   However, the primary technique for identifying CT-AGNs is through X-ray spectral fitting. This technique not only requires high-quality X-ray spectra (i.e., X-ray sources with high photon counts) of CT-AGNs but also heavily relies on absorption correction models of X-rays. 
   In the Cosmological Evolution Survey (COSMOS), \cite{2018MNRAS.480.2578L} successfully identified 67 CT-AGNs by this method from a sample of 1855 AGNs with high ($>30$) photon counts. 
   However, many CT-AGNs may still be hidden in AGNs that exhibit low photon counts or that have not been detected by X-ray instruments.
   To identify the hidden (hitherto unknown) CT-AGNs in high-redshift environments, selecting a large-scale deep X-ray survey is crucial. In this case, COSMOS is the most suitable. Establishing a series of specific tasks is also essential to carry out in COSMOS to identify these hidden CT-AGNs.
   In this work, we aim to execute and complete the first task in COSMOS, focusing on identifying CT-AGNs among X-ray-selected AGNs with low photon counts. 
   This paper is structured as follows. Sect.~\ref{sec:sample-data} describes the multiwavelength data and sample used in this work. 
   In Sect.~\ref{sec:analysis}, we analyze multiwavelength data to extract the essential physical parameters necessary for the subsequent diagnosis of CT-AGNs. We use multiwavelength approaches to diagnose CT-AGNs in Sect.~\ref{sec:diagnostic}. The discussion is presented in Sect.~\ref{sec:discussion}, followed by a summary in Sect.~\ref{sec:summary}. We adopt a concordance flat $\Lambda$-cosmology with ${\rm H_0 = 67.4 \ km\ s^{-1}\ Mpc^{-1}}$, $\Omega_{\rm m} = 0.315$, and $\Omega_\Lambda = 0.685$ \citep{2020A&A...641A...6P}.


\section{The data and sample} \label{sec:sample-data}

\subsection{The COSMOS survey}
The COSMOS is a deep field survey project that covers a relatively large area of 2 deg$^2$ \citep{2007ApJS..172....1S}, offering an extensive and unparalleled multiwavelength coverage that spans from the UV to the far-infrared (FIR) band.
In recent years, several studies have measured (or collected) multiwavelength photometric data in COSMOS and have publicly released meticulously compiled photometric catalogs, such as COSMOS2015 \citep{2016ApJS..224...24L} and COSMOS2020 \citep{2022ApJS..258...11W}. 
Furthermore, the COSMOS is also covered by several large spectroscopic projects, providing spectroscopic redshifts for a subset of the sources in this area \citep[e.g.,][]{2007ApJS..172...70L, 2009ApJS..184..218L, 2015A&A...576A..79L, 2017A&A...600A.110T}. \cite{2018ApJ...858...77H} provided the 10k DEIMOS spectroscopy catalog and publicly released the spectra covering the optical to the near-infrared (NIR).

X-ray observations in the COSMOS have provided an exceptional dataset for probing the AGNs.
These X-ray observations were obtained using the \textit{XMM-Newton} satellite, resulting in the first relatively homogeneous coverage of the field \citep{2007ApJS..172...29H,2009A&A...497..635C,2010ApJ...716..348B}. This homogeneous coverage, with a depth of approximately 60 ks across the entire COSMOS, was achieved with a total exposure time of $\sim$ 1.5 Ms.
The \textit{Chandra} satellite has observed the central 0.9 deg$^2$ of COSMOS between 2006 and 2007 \citep{2009ApJS..184..158E,2012ApJS..201...30C}, with a total exposure time of 1.8 Ms. Additionally, an additional 1.2 deg$^2$ was covered by the Chandra COSMOS Legacy Survey in 2013-2014 \citep{2016ApJ...819...62C}, with a total exposure time of 2.8 Ms.
Consequently, the entire COSMOS was observed with the \textit{Chandra} satellite, resulting in a relatively deep average coverage of approximately 160 ks, for a total of 4.6 Ms.

\subsection{X-ray data}
Our work aims to identify CT-AGNs, which are typically faint in the X-ray band. Given their faint emissions, \textit{Chandra} is the preferred instrument for detecting such sources. The X-ray data in this work were obtained from the \textit{Chandra} COSMOS Legacy Survey; specifically, the catalogs provided by \cite{2016ApJ...819...62C} and \citet[hereafter \citetalias{2016ApJ...817...34M}]{2016ApJ...817...34M}. These catalogs contain detailed information on the X-ray sources in COSMOS, including their fluxes, luminosities, counts, column densities, and other relevant properties. 
Our work mainly uses two primary parameters in the X-ray band: 0.5--7~keV aperture photometry counts and observed 2--10~keV luminosities.

\subsection{Multiwavelength photometric data}
Identifying CT-AGNs with low photon counts requires not only X-ray band data but also support from other bands. 
The dust torus surrounding CT-AGNs absorbs a significant portion of their optical-UV emission, making them challenging to detect in these wavelengths. However, their MIR emissions are less affected by this absorption, which makes MIR radiation a valuable tool for identifying these sources.
To ascertain the MIR radiation emitted by AGNs, we must construct the spectral energy distributions (SEDs) for these galaxies. This process necessitates the utilization of multiwavelength photometric data.

This work uses multiwavelength photometric data, spanning from the far-UV (1526~\AA) to the FIR (500~\micron).
Among them, the photometric data covering the wavelength range from the far-UV to MIR, totalling 40 bands, is from the COSMOS2020 catalog \citep{2022ApJS..258...11W}. These bands include the \textit{GALEX}  FUV and NUV band, CFHT U band, the HST F814W band, five Subaru HSC bands (g, r, i, z, and y), seven Subaru Suprime-Cam broad bands (B, g+, V, r$^+$, i$^+$, z$^+$, and z$^{++}$), 12 Subaru Suprime-Cam medium bands (IB427, IB464, IA484, IB505, IA527, IB574, IA624, IA679, IB709, IA738, IA767, and IB827), four VISTA  VIRCAM broad bands (Y, J, H, and $K_s$), and four Spitzer IRAC bands (ch1, ch2, ch3, and ch4). 
Moreover, the FIR photometric data is from the COSMOS2015 catalog \citep{2016ApJS..224...24L},  including MIPS 24~\micron, PACS 100~\micron, PACS 160~\micron, SPIRE 250~\micron, SPIRE 350~\micron, and SPIRE 500~\micron.

\subsection{Optical-NIR spectra}
Accurate optical emission line information, such as the [O~III]$\lambda$5007 luminosity, is crucial in identifying CT-AGNs using multiwavelength approaches.
However, considering that most of the sources are at high redshifts, their optical emission lines may be shifted into the NIR band in the observer's frame.
Therefore, we have only used the optical-NIR spectra provided by \cite{2018ApJ...858...77H}.

\subsection{Sample selection}
The starting point of the analysis is the X-ray sources in the \textit{Chandra} COSMOS Legacy Survey \citep[e.g.,][and \citetalias{2016ApJ...817...34M}]{2016ApJ...819...62C}. Among these X-ray sources, the majority are believed to be AGNs, as is stated in \cite{2019ApJ...872..168S}. Based on the X-ray sources catalog provided by \cite{2016ApJ...819...62C} and \citetalias{2016ApJ...817...34M}, we further selected our AGN sample using the following three criteria:
\begin{enumerate}[(1)]
        \item The X-ray source is extragalactic, and has a reliable optical counterpart and spectroscopic or photometric redshift in \citetalias{2016ApJ...817...34M}.
        \item Sources are characterized by a net count of fewer than 30 within the entire 0.5--7 keV band. 
        \item The source is required to have multiwavelength photometric data, which must include four Spitzer IRAC bands and MIPS 24~\micron.
\end{enumerate}
The first criterion guarantees that each source is an AGN, as is stated in \cite{2019ApJ...872..168S}. 
The second criterion is designed to ensure complementarity with the work conducted by \cite{2018MNRAS.480.2578L}, as accurate column density measurements (at a higher than 90 percent confidence level) for low photon count sources using X-ray spectrum fitting is very difficult.
The third criterion ensures that each source has sufficient photometric data for SED fitting. This criterion is crucial for obtaining accurate measurements of the physical properties of the sources, including the AGN's contribution to the MIR band.

We successfully constructed a sample of 440 AGNs based on the above three criteria. Within this sample, 72 sources have been detected in the PACS 100~\micron~band, and 113 sources have optical-NIR spectra available.

\subsection{Redshift selection and correction of luminosity }
Redshifts are a fundamental parameter in studying galaxies and AGNs, and are crucial for revealing their intrinsic properties. Therefore, using accurate and reliable redshifts is imperative for our AGN sample.

We collected the latest spectral redshifts from the \citetalias{2016ApJ...817...34M} catalog and \cite{2018ApJ...858...77H} catalog. We prioritized adopting the spectral redshift of a source whenever its redshift quality flag was at least 1.5. 
If the quality flag of a source is less than 1.5 or if it is absent from the spectral redshift, we adopt its photometric redshift.
We have also collected the latest photometric redshifts from the \citetalias{2016ApJ...817...34M} catalog.
Moreover, we have also remeasured the photometric redshifts of these sources using SED fitting. The reason is that we found poor results for a subset of sources during SED fitting. The likely cause of these poor fitting results is the inaccuracy of the photometric redshifts provided by the \citetalias{2016ApJ...817...34M} catalog for these sources. In Appendix~\ref{sec:my-photoz}, we present two examples that demonstrate significant improvements in their SED fitting when utilizing our photometric redshifts. Therefore, we have used the photometric redshifts from the \citetalias{2016ApJ...817...34M} catalog or the ones we derived. The photometric redshifts were selected based on the following two criteria:
\begin{enumerate}[(1)]
        \item For a source, if the photometric redshift provided by the \citetalias{2016ApJ...817...34M} catalog is in agreement with the one we derived ($\Delta z$/(1 + $z_{M16}$) $<$ 0.15), we prefer to adopt the photometric redshift from the \citetalias{2016ApJ...817...34M} catalog. 
        \item If the photometric redshift of a source provided by \citetalias{2016ApJ...817...34M} catalog does not agree with the one we derived ($\Delta z$/(1 + $z_{M16}$) $>$ 0.15), we adopt the photometric redshift that can make its SED fitting better. 
\end{enumerate}
In our sample, 236 sources use a spectral redshift, and 204 use a photometric redshift. Among them, 38 AGNs use our photometric redshifts. Cols. 4--6 of Table~\ref{Tab:summary} show the redshifts, redshift types, and the references of redshift for our sample.

\begin{figure}
        \includegraphics[width=\linewidth]{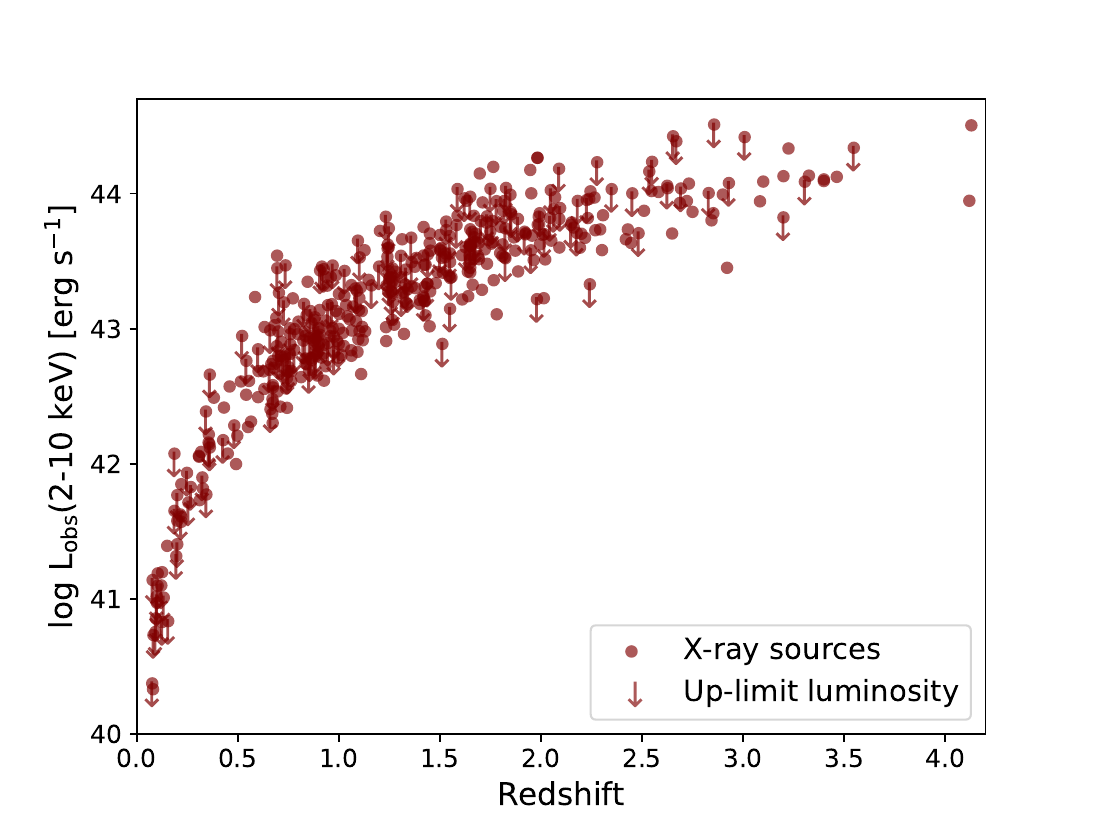}
        \caption{Rest-frame 2--10~keV luminosity vs. redshift. The solid red circles represent our sample of AGNs. The downward arrows indicate that the X-ray luminosities for these AGNs are based on their upper limits.}
        \label{fig:z-Lx}
\end{figure}

Since we have used more accurate redshifts relative to the \citetalias{2016ApJ...817...34M} catalog, the observed X-ray luminosities of these sources need to be corrected.
The newly corrected X-ray luminosity is derived by
\begin{equation}
        \frac{L_{new}}{L_{old}}= \frac{R(z_{new})^2 (1+z_{new})^{\Gamma-2}}{R(z_{old})^2 (1+z_{old})^{\Gamma-2}},
        \label{equ:cor}
\end{equation}
where $L_{new}$ and $L_{old}$ are the newly corrected X-ray luminosity and observed X-ray luminosity in the \citetalias{2016ApJ...817...34M} catalog, respectively,  $z_{new}$ and $z_{new}$ are our redshift and the redshift from \citetalias{2016ApJ...817...34M} catalog, respectively, $R(z)$ is the luminosity distance, and $\Gamma$ is the photon index of an X-ray spectrum assuming 1.4\footnote{The observed X-ray luminosities provided by \citetalias{2016ApJ...817...34M} also assumed that their photon indexes are 1.4. }. Col. 7 of Table~\ref{Tab:summary} lists observed 2--10~keV luminosities for our sample. Figure~\ref{fig:z-Lx} shows the distribution of observed 2--10~keV luminosities with redshifts in our sample.

\section{Analysis of  multiwavelength data}\label{sec:analysis}
\subsection{Analysis of SEDs}
The multiband SED analysis of AGNs is a widely used method that not only provides detailed information about AGNs but also reveals the properties of their host galaxies \citep[e.g.,][]{2020MNRAS.492.1887G}, such as AGN luminosity, rest-frame 6~\micron~luminosity for the AGN, stellar mass (M$_\star$), and the star formation rate (SFR). In this section, we use the SED fitting \textit{Code Investigating GALaxy Emission} \cite[CIGALE; V2022.1;][]{2019A&A...622A.103B}, an open \textit{Python} code containing the template of galaxies and AGNs, to analyze our sample. 

\begin{figure*}
        \includegraphics[width=0.49\linewidth]{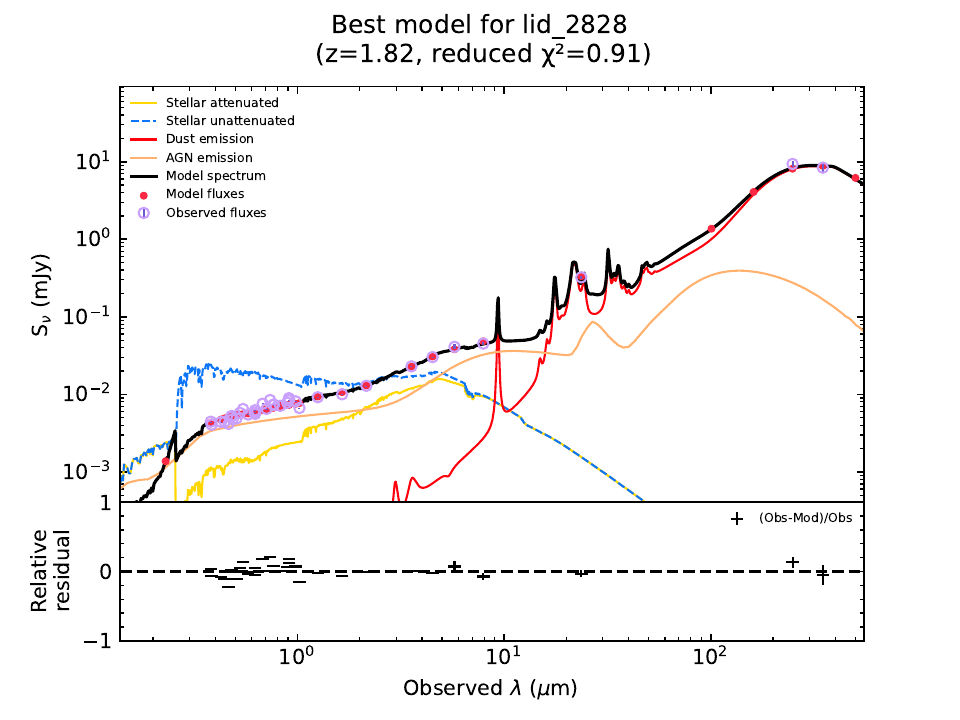}
        \includegraphics[width=0.49\linewidth]{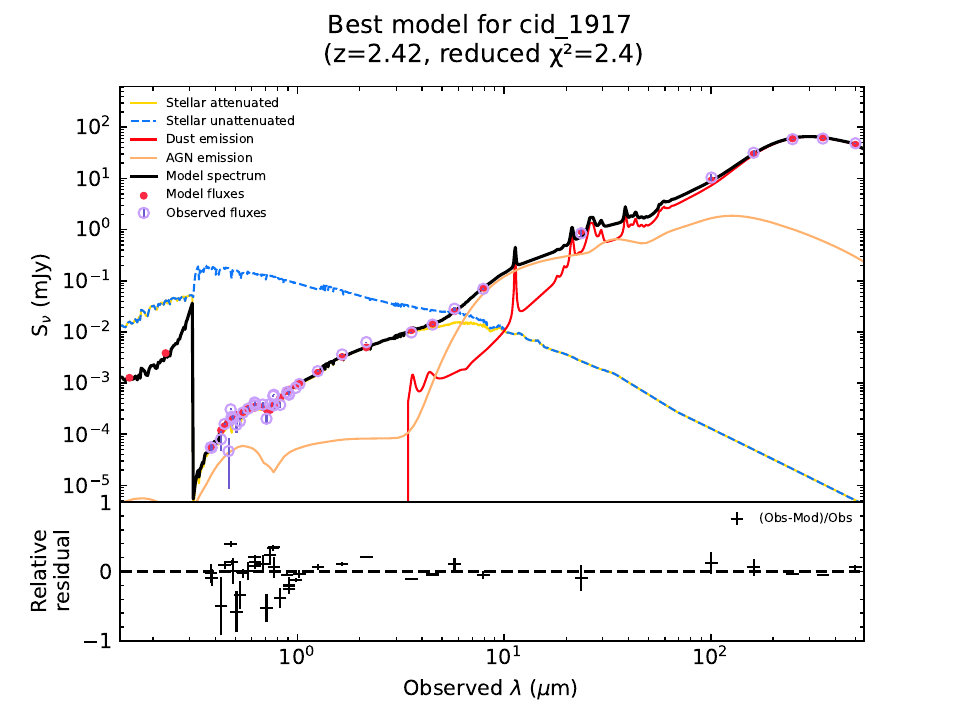}
        \caption{Examples of best-fitting SEDs for type 1 and 2 AGNs. The solid black line indicates the best-fitting model. The dashed blue, solid gold, and solid red lines represent unattenuated stellar, attenuated stellar, and dust emission, respectively. The solid apricot line indicates AGN emission. The solid strawberry and pastel purple open circles represent model and observed fluxes, respectively. The lower panel indicates the residual of the best fitting.}  
        \label{fig:SED-example}
\end{figure*}

Given that all the sources in our sample are X-ray AGNs, we have used templates that account for the galaxy and AGN characteristics to achieve the most accurate fit for their SEDs. The templates for the galaxy in CIGALE are created by integrating four distinct modules, including the star formation history (SFH), the single stellar population (SSP) model \citep{2003MNRAS.344.1000B}, the dust attenuation \citep[DA;][]{2000ApJ...533..682C}, and the dust emission \citep[DE;][]{2007ApJ...657..810D,2014ApJ...784...83D}. The AGN component within our analysis is a module developed by \cite{2006MNRAS.366..767F}. The modules and parameters used for SED fitting are summarized in Table~\ref{Tab:SED-module}.
Due to the vast parameter space required, quickly constructing a model on the computer is not feasible. We adopted an iterative methodology to expedite the acquisition of the best-fitting SEDs. Figure~\ref{fig:SED-example} presents the two best-fitting SEDs. The left panel shows an SED of an optical type 1 AGN, and the right one presents an SED of a type 2 AGN. 
To estimate the reliability of the fit AGN components, we refit their SEDs using only the galaxy templates. If the fit with the galaxy+AGN template results in a significantly better goodness-of-fit statistic (i.e., lower $\chi^2$) compared to the galaxy-only template, then it suggests that the fit AGN components can indeed represent the contribution of the actual AGNs in their SEDs.
We used the Bayesian information criterion (BIC) to quantify the AGN components' significance or confidence (please refer to Appendix~\ref{sec:quantifying} for details).

\begin{table*}
        \caption{Module assumptions for SED fitting.}             
        \label{Tab:SED-module}      
        \centering          
        \resizebox{\hsize}{!}
        {
        \begin{tabular}{l|c|c|c}     
                \hline\hline       
                Component & Module & Parameter & Value \\
                \multicolumn{1}{c|}{(1)}&(2)&(3)&(4)\\
                \hline                    
                \multirow{13}{*}{Galaxy}&\multirow{5}{*}{SFH(delayed)}&tau\_main (Myr)&20 -- 8000 (in steps of 10)\\
                &&age\_main (Myr)&200 -- 13000 (in steps of 10)\\
                &&tau\_burst (Myr)&10 -- 200(in steps of 1)\\
                &&age\_burst (Myr)&10 -- 200(in steps of 1)\\
                &&f\_burst &0, 0.0001, 0.0005, 0.001, 0.005, 0.01, 0.05, 0.1, 0.15,0.20, 0.25, 0.3, 0.40, 0.50\\
                \cline{2-4}
                &\multirow{2}{*}{SSP(BC03)}&imf&1 (Chabrier)\\
                &&metallicity &0.02\\
                \cline{2-4}
                &DA&\multirow{2}{*}{E\_BV\_nebular (mag)}&0.005, 0.01, 0.025, 0.05, 0.075, 0.10, 0.15,0.20, 0.25, \\
                &(dustatt\_calzleit)&&0.30, 0.35, 0.40, 0.45, 0.50, 0.55, 0.60\\ 
                \cline{2-4}
                &\multirow{4}{*}{DE(dl2014)}&qpah& 1.12, 1.77, 2.50, 3.19\\
                &&umin &5.0, 6.0, 7.0, 8.0, 10.0, 12.0, 15.0, 17.0, 20.0, 25.0\\
                &&alpha &2.0, 2.1, 2.2, 2.3, 2.4, 2.5, 2.6, 2.7, 2.8\\
                &&gamma &0.02\\
                \hline
                \multirow{8}{*}{AGN}&\multirow{8}{*}{AGN(Fritz2006)}&r\_ratio&10, 30, 60, 100, 150\\
                &&tau&0.1, 0.3, 0.6, 1.0, 2.0, 3.0\\
                &&beta&-1.00, -0.75, -0.50, -0.25, 0.00\\
                &&gamma&0.0, 2.0, 4.0, 6.0\\
                &&opening\_angle&60, 100, 140\\
                &&psy&0.001, 10.1, 20.1, 30.1, 40.1, 50.1, 60.1, 70.1, 80.1, 89.99\\
                &&\multirow{2}{*}{fracAGN}&0.0, 0.05, 0.1, 0.15, 0.2, 0.25, 0.3, 0.35, 0.4, 0.45, 0.5, 0.55, 0.6, 0.65, 0.7, 0.75, \\
                &&&0.8, 0.85, 0.9,
                0.95, 0.99\\
                \hline                  
        \end{tabular}
}
\end{table*}

Through SED fitting, we secured the best-fitting SED for each source and derived its corresponding physical parameters. Even though the sources in our sample are all AGNs, some sources cannot be fit with AGN components through SED modeling because the AGN emission of these sources is weak. There are 329 AGNs with fit AGN components. Cols. 8--9 of Table~\ref{Tab:summary} are rest-frame 6~\micron~luminosity for the AGN and the AGN components' confidence. 
Col. 10 of Table~\ref{Tab:summary} presents M$_\star$, the derived outcome of SED fitting. However, since the SFR derived from SED fitting relies on the chosen SFH model, we used the calibration method proposed by \cite{2005ApJ...625...23B} to estimate the SFR using UV and IR luminosities. This calibration was scaled to the \cite{2003PASP..115..763C} initial mass function for accuracy. Specifically, the SFR was calculated using the formula
\begin{equation}
        \label{equ:SFR}
        \mathrm{SFR(M_\odot\ yr^{-1})}=1.09\times10^{-9} (3.3\mathrm{L_{UV}}+\mathrm{L_{IR}}),
\end{equation}
where $\mathrm{L_{UV}}=\nu \mathrm{L}_\nu$ is an estimation of the integrated 1216 -- 3000~\AA~rest-frame UV luminosity, and $\mathrm{L_{IR}}$ is the 8 -- 1000~\micron~rest-frame IR luminosity. Both $\mathrm{L_{UV}}$ and $\mathrm{L_{IR}}$ are in units of $ \mathrm{L}_\odot$. Col. 11 of Table~\ref{Tab:summary} lists the SFR of each source.

\subsection{Analysis of optical-NIR spectra}
The AGN spectra contain a wealth of information that can provide us with their various properties. Spectral fitting is the most widely used and effective technique for extracting information from AGN spectra. Using the spectral fitting technique, we can decompose the observed AGN spectrum into its various contributing components, including the contribution from the host galaxy, the continuous emission of the AGN, and its distinct emission lines.
This section uses the publicly available \textit{Python} code \textit{PyQSOFit} \citep{2018ascl.soft09008G, 2019ApJS..241...34S} for spectral fitting. By fitting the optical-NIR spectra in our sample, we can extract information on emission lines for some sources.  Cols. 12--13 of Table~\ref{Tab:summary} present the luminosities and signal-to-noise ratios (S/Ns) of the [O~III]$\lambda~5007$ emission line for 21 AGNs. Cols. 14--15 of Table~\ref{Tab:summary} present the luminosities and S/Ns of the  [O~II]$\lambda~3727$ emission line for 56 AGNs. There are nine AGNs with the [Ne~V]$\lambda~3426$ emission line, and their luminosities and S/Ns are listed in Cols. 16--17 of Table~\ref{Tab:summary}. 

\section{Multiwavelength diagnostic CT-AGNs}\label{sec:diagnostic}
\subsection{MIR diagnostics}

According to the unified model of AGNs, the emission from both the corona and the dust torus can be good tracers of the accretion disk's emission. Consequently, there should be a strong correlation between the X-ray and MIR emissions of AGNs. Some studies have investigated the relationship between X-ray and MIR emissions in radio-quiet AGNs \citep[e.g.,][]{2009ApJ...693..447F,2015MNRAS.449.1422M,2015ApJ...807..129S,2017ApJ...837..145C,2019ApJ...886..125E,2021A&A...651A..91E}.
Due to the low optical depth in the MIR band, even the MIR radiation from CT-AGNs is not subject to significant suppression. The X-ray radiation from CT-AGNs is heavily absorbed, resulting in observed X-ray luminosities that are lower than expected for these relationships. Therefore, we used these relationships to identify sources where the X-ray radiation is heavily absorbed.

\begin{figure}
        \includegraphics[width=\linewidth]{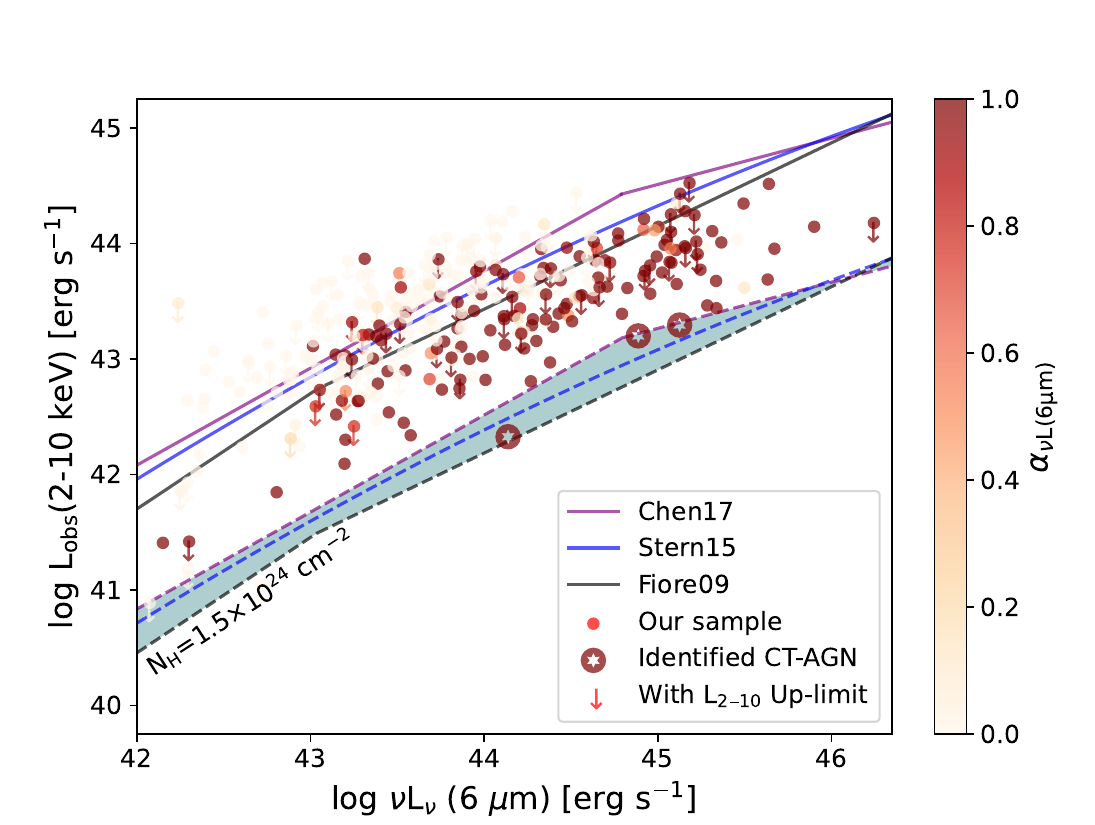}
        \caption{Observed X-ray luminosity in the rest-frame 2--10 keV band as a function of the 6~\micron~luminosity for fit AGN component. The solid purple, blue, and black lines represent the relation for \cite{2017ApJ...837..145C}, \cite{2015ApJ...807..129S}, and \cite{2009ApJ...693..447F}, respectively. The dashed lines indicate the same relationships but where the X-ray luminosities are absorbed by a column density of $\mathrm{N_H}=1.5\times 10^{24}\ \mathrm{cm}^{-2}$. The shaded area represents the column density of absorbed gas is $1.5\times 10^{24}\ \mathrm{cm}^{-2}$. The solid circles are 329 AGNs with a fit AGN component. The different colors represent  the confidence coefficient of the 6~\micron~luminosity for a fit AGN component (or the fit AGN components). The cutting stars within the large circles represent CT-AGNs identified using the relationship between MIR and X-ray luminosities.}
        \label{fig:midIR-diagnostics}
\end{figure}

Figure~\ref{fig:midIR-diagnostics} shows the observed X-ray luminosity in the rest-frame 2--10 keV band as a function of the 6~\micron~luminosity for the fit AGN component. The solid circles represent our sample. The downward arrows represent the upper limit of X-ray luminosities. The different colors represent the confidence of the fit AGN components. The solid purple, blue, and black lines represent the relation for \cite{2017ApJ...837..145C}, \cite{2015ApJ...807..129S}, and \cite{2009ApJ...693..447F}, respectively. 
Assuming that the intrinsic spectra of AGNs can be described by a model (\textit{phabs $\cdot$ powerlaw + constant $\cdot$ powerlaw + pexmon}, please refer to Appendix~\ref{sec:absorb} for more detailed information) in \textit{Xspec} within the 2--10 keV energy range, the dashed lines represent the same relationships between MIR and X-ray luminosities after the X-ray spectra have been absorbed by gas with a column density of $\mathrm{N_H}=1.5\times 10^{24}\ \mathrm{cm}^{-2}$. 
This means that the sources within or below the shaded area in Fig.~\ref{fig:midIR-diagnostics} should be classified as CT-AGNs. 
According to Fig.~\ref{fig:midIR-diagnostics}, five sources fall within or below the shaded area, indicating the possibility that they could be CT-AGNs.
Given that the fit AGN components of XID \textit{cid\_3441} and\textit{ lid\_2104} have a low confidence coefficient ($\alpha_{AGN} < 0.95$), their SED fitting derived AGN components may be unreliable, making it difficult to diagnose these sources as CT-AGNs using this method. Conversely, the AGN components in the remaining three sources exhibit a high confidence coefficient ($\alpha_{AGN} > 0.95$), supporting their diagnosis as CT-AGNs. Table~\ref{table:new-CT-AGN} lists their XIDs.

\subsection{Optical emission line diagnostics}\label{sec:optical-emission}

Most CT-AGNs are categorized as optical type-2 AGNs, indicating that their optical spectra typically lack broad emission lines. According to the unified model of AGNs, the broad-line region in CT-AGNs is obscured by the dust torus, while the narrow-line region (NLR) remains unobscured. The radiation from the accretion disk illuminates the clouds in the NLR, ionizing them and causing them to emit high-ionization narrow emission lines. Therefore, as long as the AGN unified model holds, the high-ionization narrow emission lines can be a reliable tracer of the radiation from the accretion disk. It is expected that there are some correlations between the luminosities of the high-ionization narrow emission lines and X-rays, and some works have investigated these relationships \citep[e.g.,][]{2005ApJ...634..161H, 2006A&A...455..173P, 2015MNRAS.454.3622B, 2015ApJ...815....1U, 2017ApJ...846..102M, 2010A&A...519A..92G}. This section uses the high-ionization narrow emission lines as a diagnostic tool to identify CT-AGNs.

\begin{figure}
        \includegraphics[width=\linewidth]{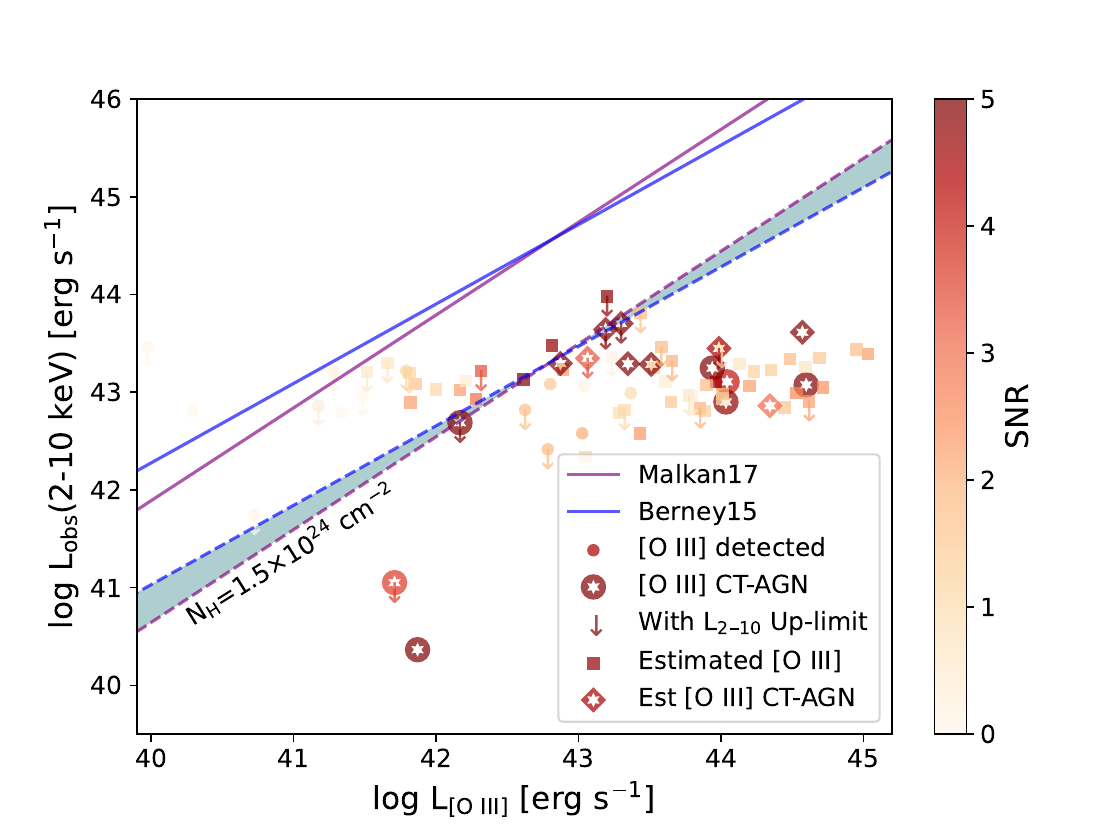}
        \caption{Observed X-ray luminosity in the rest-frame 2--10 keV band as a function of the  [O~III]$\lambda~5007$ luminosity. The solid purple and blue lines represent the relation for \cite{2017ApJ...846..102M} and \cite{2015MNRAS.454.3622B}. The dashed lines indicate the same relationships but where the X-ray luminosities are absorbed by a column density of $\mathrm{N_H}=1.5\times 10^{24}\ \mathrm{cm}^{-2}$. The shaded area represents that the column density of absorbed gas is $1.5\times 10^{24}\ \mathrm{cm}^{-2}$. The solid circles are 21 AGNs with [O~III]$\lambda~5007$ detected. The different colors represent the S/N of the emission line. The cutting stars within the large circles represent CT-AGNs identified using the relationship between [O~III]$\lambda~5007$ and X-ray luminosities. The solid squares are 56 AGNs with [O~II]$\lambda~3727$ detected. The cutting stars within the large squares indicate CT-AGNs identified utilizing the relationship between estimated [O~III]$\lambda~5007$ luminosities and X-ray luminosities.}
        \label{fig:OIII-diagnostics}
\end{figure}

The [O~III]$\lambda~5007$ emission line is a prominent feature in the spectra of AGNs, and it can trace the emission from the accretion disk well. Therefore, it has a close relationship with X-ray emission. Some studies use the observation of strong [O~III]$\lambda~5007$ emission lines with weak or absent X-ray radiation to diagnose CT-AGNs \citep[e.g.,][]{2020ApJ...905L...2C,2022ApJ...936..162S}. Similar methods can be used to diagnose CT-AGNs at $z\lesssim 1$. Figure~\ref{fig:OIII-diagnostics} presents the observed X-ray luminosity in the rest-frame 2--10 keV band as a function of the [O~III]$\lambda~5007$ luminosity. The solid purple and blue lines represent the relations between \cite{2017ApJ...846..102M} and \cite{2015MNRAS.454.3622B}. The dashed lines indicate the same relationships after being absorbed by $\mathrm{N_H}=1.5\times 10^{24}\ \mathrm{cm}^{-2}$ gas. The solid circles represent the AGNs with [O~III]$\lambda~5007$ detected. The downward arrows represent the upper limit of X-ray luminosities. The different colors represent the S/N of the emission line. 
In Fig.~\ref{fig:OIII-diagnostics}, a total of 13 solid circles are situated either within or below the shaded area. Of these, seven sources (\textit{cid\_1092, cid\_1368, lid\_3021, lid\_3603, lid\_375, lid\_3846} and \textit{lid\_3901}) exhibit an S/N for the [O~III]$\lambda~5007$ emission line that exceeds the threshold of 3.
Based on our diagnosis, these seven AGNs with S/N $>$3 can be diagnosed as CT-AGNs.

If the sources are at $z>1$, their [O~III]$\lambda~5007$ emission lines will shift beyond the range of their optical-NIR spectra, making it impossible to directly obtain their [O~III]$\lambda~5007$ luminosities through spectral fitting. However, we can still indirectly estimate their [O~III]$\lambda~5007$ luminosities by analyzing shorter-wavelength emission lines with a known ratio to the [O~III]$\lambda~5007$ emission line. Given that a considerable number of sources have detections of the [O~II]$\lambda~3727$ emission line in their optical-NIR spectra, we intend to use the [O~II]$\lambda~3727$ luminosities as a basis for estimating the [O~III]$\lambda~5007$ luminosities. The classic [O~II]/[O~III] ratio typically falls within the range of 0.1 to 0.6 when the AGN luminosities are between $10^{43}$ and  $10^{45}\ \mathrm{erg\cdot s^{-1}}$ \citep[][in Sect.~11.4]{2013peag.book.....N}. Assuming that the upper limit of the [O~II]/[O~III] ratio in AGNs is taken as 0.6, the lower limit of the [O~III]$\lambda~5007$ luminosities can be estimated using 
\begin{equation}
        \mathrm{L_{[O~III]}} \geqslant \frac{\mathrm{L_{[O~II]}}}{\mathrm{[O~II]/[O~III]}},
\end{equation}
where $\mathrm{L_{[O~III]}}$ is the luminosity of the [O~III]$\lambda~5007$ emission line, and $\mathrm{L_{[O~II]}}$ is the known luminosity of the [O~II]$\lambda~3727$ emission line. 
Similarly, we also used the relationship between  [O~III]$\lambda~5007$ and X-ray luminosity to diagnose CT-AGNs. In Fig.~\ref{fig:OIII-diagnostics}, the solid squares represent the AGNs with an estimated [O~III]$\lambda~5007$. A total of 43 solid squares are located either within or below the shaded area. However, ten sources have S/N $>$3, and can be diagnosed as CT-AGNs. Their XIDs are also listed in Table~\ref{table:new-CT-AGN}.

\begin{figure}
        \includegraphics[width=\linewidth]{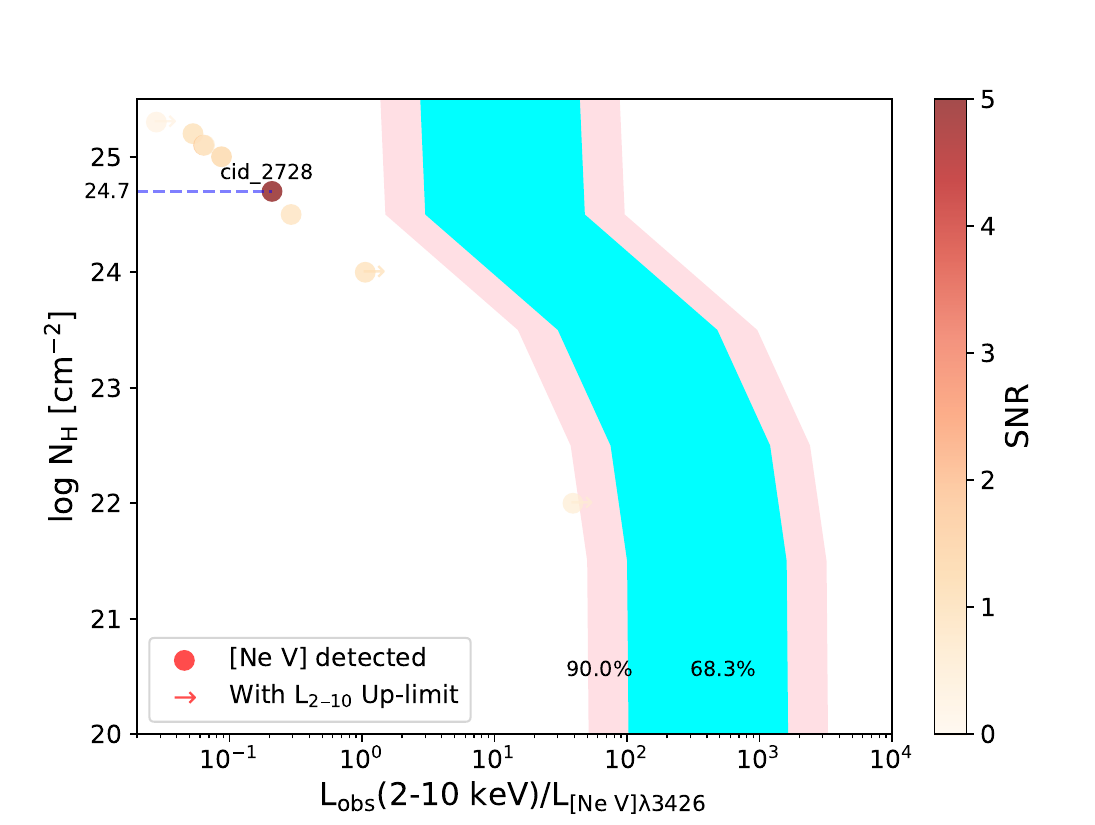}
        \caption{Observed 2–10 keV to [Ne~V]$\lambda~3426$ luminosity ratio as a function of $\mathrm{N_H}$. The shaded cyan and pink regions correspond to 68.3\% and 90\% around the $<$X/[Ne~V]$>$ ratio \citep{2010A&A...519A..92G}.}
        \label{fig:NeV-diagnostics}
\end{figure}

Like the [O~III]$\lambda~5007$ emission line, 
the [Ne~V]$\lambda~3426$ is also a high-ionization emission line and can also trace the emission from the accretion disk. Thus, the relationship between [Ne~V]$\lambda~3426$ and X-ray luminosity is also used to diagnose CT-AGNs. \cite{2010A&A...519A..92G} provided a diagnostic based on the $<$X/[Ne~V]$>$ ratio. Figure~\ref{fig:NeV-diagnostics} shows the $\mathrm{L_X^{obs} /L_{[NeV]}}$ ratio as a function of $\mathrm{N_H}$. Based on the information in Fig.~\ref{fig:NeV-diagnostics}, we can derive that the column densities of seven sources are more than  $10^{24} \ \mathrm{cm}^{-2}$ with a 90\% probability. Given that the S/N of [Ne~V]$\lambda~3426$ emission line is greater than 3 for only one source, only the source of XID \textit{cid\_2728} can be diagnosed as a CT-AGN  based on this method. 

In short, we have successfully identified 18 CT-AGNs with low photon counts using the multiwavelength approaches. The summary of the diagnostics is listed in Table~\ref{table:new-CT-AGN}.

\begin{table}
        \caption{Diagnostic CT-AGNs.} \label{table:new-CT-AGN}
        \centering
        \resizebox{\hsize}{!}
        {
        \begin{tabular}{l l c c c c}
                \hline\hline
                XID & z & \multicolumn{4}{c}{Diagnostics} \\ 
                \cline{3-6}
                &&MIR&[O~III]&Estimated [O~III]&[Ne~V]\\
                (1)&(2)&(3)&(4)&(5)&(6)\\
                \hline
                cid\_1092 & 0.361 & $\cdots$ & CT & $\cdots$ & $\cdots$\\ 
                cid\_1244 & 1.028 & $\cdots$ & $\cdots$ & CT & $\cdots$\\ 
                cid\_1368 & 0.098 & $\cdots$ & CT & $\cdots$ & $\cdots$\\ 
                cid\_1750 & 1.061 & $\cdots$ & $\cdots$ & CT & $\cdots$\\ 
                cid\_2462 & 1.548 & $\cdots$ & $\cdots$ & CT & $\cdots$\\ 
                cid\_2728 & 1.506 & Not CT & $\cdots$ & CT & CT\\ 
                cid\_3570 & 1.244 & Not CT & $\cdots$ & CT & $\cdots$\\ 
                cid\_3714 & 1.265 & $\cdots$ & $\cdots$ & CT & $\cdots$\\ 
                cid\_437 & 1.242 & $\cdots$ & $\cdots$ & CT & $\cdots$\\ 
                lid\_2071 & 1.329 & $\cdots$ & $\cdots$ & CT & $\cdots$\\ 
                lid\_3021 & 0.881 & $\cdots$ & CT & CT & $\cdots$\\ 
                lid\_3603 & 0.772 & Not CT & CT & May CT & $\cdots$\\ 
                lid\_3702 & 1.400 & CT & $\cdots$ & $\cdots$ & $\cdots$\\ 
                lid\_375 & 0.080 & $\cdots$ & CT & $\cdots$ & $\cdots$\\ 
                lid\_3846 & 0.800 & $\cdots$ & CT & May CT & $\cdots$\\ 
                lid\_3901 & 0.881 & Not CT & CT & May CT & $\cdots$\\ 
                lid\_673 & 1.452 & CT & $\cdots$ & CT & $\cdots$\\ 
                lid\_967 & 0.672 & CT & $\cdots$ & $\cdots$ & $\cdots$\\ 
                \hline
        \end{tabular}
}
        \tablefoot{(1) and (2) are XID and redshift. (3), (4), (5) and (6) are diagnostics of MIR and optical emission lines.}
\end{table}

\section{Discussion}\label{sec:discussion}

\subsection{The reasons for the discrepancy in the diagnosis of CT-AGNs}
In the previous section, the diagnosis of CT-AGNs relied on two methods. Among the 18 diagnosed CT-AGNs, only one source fulfilled the diagnostic criteria for CT-AGNs with the two methods. The majority of CT-AGNs can only be diagnosed through a single approach. Either their data is absent, or their confidence level in the data is insufficient for a reliable diagnosis using another method. Moreover, there are discrepancies in the results of four AGNs, specifically \textit{cid\_2728}, \textit{cid\_3570}, \textit{lid\_3603}, and \textit{lid\_3901}, when diagnosed through two methods: the emission line diagnostics suggest they are CT-AGNs; the MIR diagnostics indicate that they are heavily absorbed but do not reach the level of CT-AGNs.
Among them, three sources have reliable detections of high-ionization emission lines. Only one source, XID \textit{cid\_3570}, lacks reliable detections of high-ionization emission lines.

The potential reasons for diagnostic discrepancies among the three sources with high-ionization emission lines may be the variability of AGNs. Using X-ray, MIR photometric, and spectral data observed at different epochs can lead to inconsistencies in the classification of these AGNs for different diagnostic methods due to their data potential for significant variability. This suggests that they may have been CT-AGNs at some epochs.
Therefore, we regard them as potential CT-AGNs, acknowledging the need for further investigation to confirm their classification.

The diagnosis of the source with XID \textit{cid\_3570} as a CT-AGN is based on the estimation of [O~III]$\lambda~5007$ luminosity using the observed [O~II]$\lambda~3727$ emission line. Thus, there is a risk of overestimating [O~III]$\lambda~5007$ luminosity, leading to a misdiagnosis as a CT-AGN, but this likelihood is very low. A more exhaustive discussion regarding this matter is presented in the following section. Therefore, the reasons for the discrepancy in the diagnosis of the source (\textit{cid\_3570}) may be similar to those of the other three sources. This source is still considered a CT-AGN.

\subsection{Whether the estimated [O~III] luminosity will be misdiagnosed as a CT-AGN}

The [O~III]$\lambda~5007$ is a high-ionization emission line, which is excited predominantly by an AGN. However, the [O~II]$\lambda~3727$ is a low-ionization emission line, which can be excited by star formation and nuclear activity. 
The emission of [O~III]$\lambda~5007$ and [O~II]$\lambda~3727$ from AGNs should show a specific linear correlation.
\cite{2019ApJ...882...89Z} demonstrate through calculations that, under plausible conditions, the NLR of high-ionization or highly accreting AGNs emits a highly restricted ratio of [O~II]/[O~III] $\approx$ 0.1.
In fact, the observed [O~II]$\lambda~3727$ radiation includes a component originating from their host galaxies. The AGN sample analyzed by \cite{2018ApJ...861L...2T} suggests that the contribution of the AGN to the [O~II]$\lambda~3727$ is within the range of approximately 20\%--100\%. Therefore, the ratio of [O~II]/[O~III] is at least 0.1, and estimated to be 0.1--0.5.
Considering the potential high SFRs in the host galaxies of the CT-AGNs, the lower limits of the luminosities for the [O~III]$\lambda~5007$ emission lines were estimated using a higher ratio of 0.6 in Sect.~\ref{sec:optical-emission}. The estimated lower limits are likely to be lower than the actual values, as the ratio of 0.6 is the lowest threshold chosen for the conservative estimation of their luminosities. In our sample, both [O~III] $\lambda~5007$ and [O~II]$\lambda~3727$ emission lines are observed in the spectra of some sources. The lower luminosity limits, which were estimated from the [O~II]$\lambda~3727$ luminosities, are indeed found to be lower than the observed [O~III] $\lambda~5007$ luminosities when there is a high S/N.

The method of diagnosing CT-AGNs based on [O~III]$\lambda~5007$ emission lines relies on a lower-than-expected X-ray to [O~III] ratio. When the [O~III]$\lambda~5007$ luminosities were underestimated, the X-ray to [O~III] ratios appeared higher than they were. This can lead to some CT-AGNs being misdiagnosed as Compton-thin AGNs because the elevated ratio might suggest less X-ray absorption than is truly present. In this work, CT-AGNs will not be misdiagnosed by using the estimated [O~III]$\lambda~5007$ luminosities; however, there is a possibility that some CT-AGNs are missed.

\subsection{Analysis of CT-AGN fraction and biases in identification}

In this work, we have identified 18 CT-AGNs, slightly increasing their fraction among the AGNs in COSMOS. The fraction of CT-AGNs still falls short of the theoretical expectation for the CXB. Our sample focuses on AGNs with low photon counts, which are expected to have a higher fraction of CT-AGNs than samples with high photon counts. However, the fraction of CT-AGNs in our sample is $\sim$ 4.1\% (18/440), which is not significantly higher than the fraction (67/1855) of the sample for \cite{2018MNRAS.480.2578L}. The selection bias of CT-AGNs may cause this to happen.

In this work, identifying CT-AGNs relies on two primary indicators: reliable AGN components in their SEDs and detecting trustworthy high-ionization emission lines in their spectra. The AGNs in our sample that meet these two indicators are inherently biased. Additionally, considering that some sources may have underestimated [O~III]$\lambda~5007$ luminosities by using the [O~II]$\lambda~3727$ luminosities, this could result in missed potential CT-AGNs during the selection process.
Therefore, our methodology inherently contains notable biases when selecting CT-AGNs, potentially leading to missed CT-AGNs in our sample.
To further increase the fraction of CT-AGNs in this field, several additional works may be necessary to improve their identification in the future.

\subsection{The properties of CT-AGN host galaxies}
\begin{figure}
        \includegraphics[width=\linewidth]{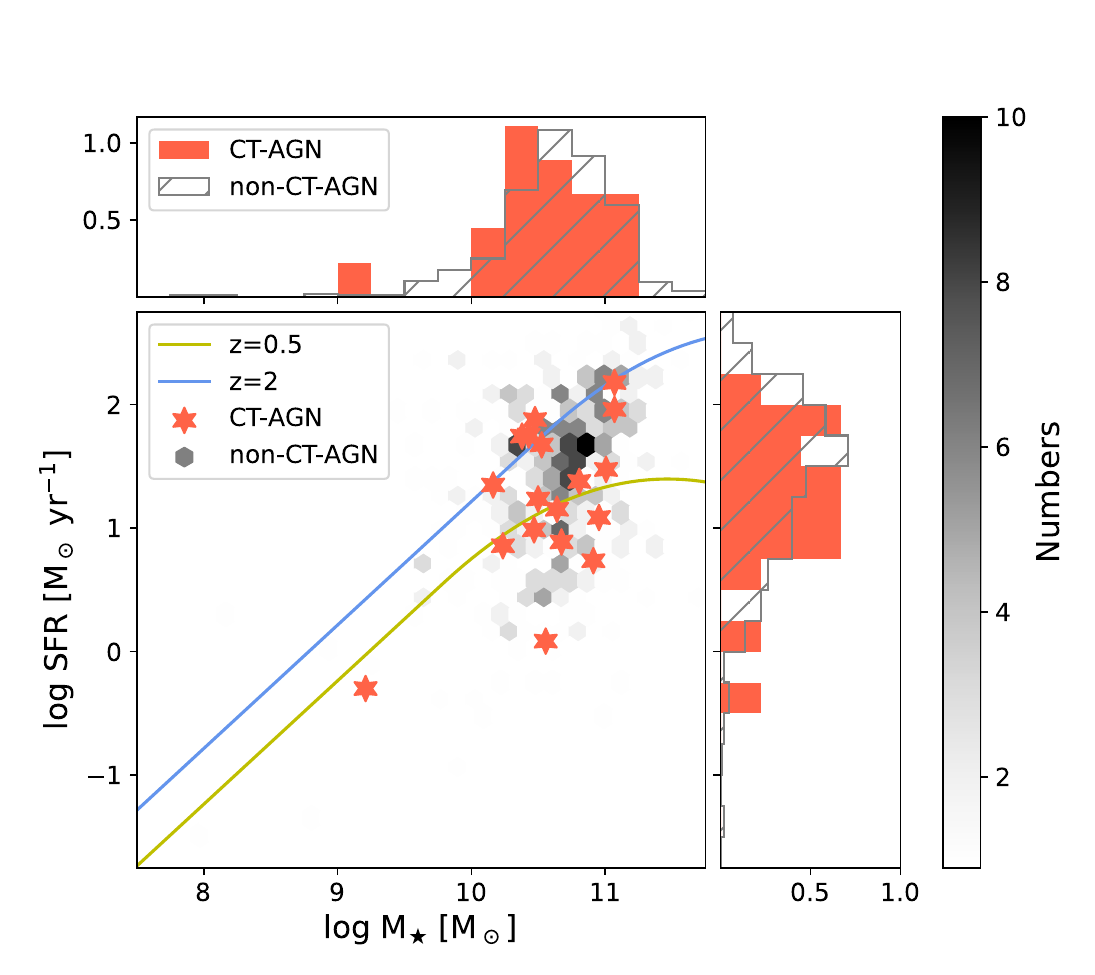}
        \caption{SFR vs. M$_\star$ distribution for our sample. For comparison, the data points of 18 CT-AGNs (solid red hexagonal stars) and non-CT-AGNs (solid hexagonal) are plotted. The yellow and blue lines show the main sequences of star formation at z$\sim$0.5 and z$\sim$2, respectively \citep{2015A&A...575A..74S}.}
        \label{fig:SFR-M}
\end{figure}

CT-AGNs are often believed to be an early phase of the evolutionary scenario for AGNs \citep[e.g.,][]{2023PASP..135a4102G}. Numerous studies have demonstrated that the evolution of AGNs is closely related to their surrounding environment, meaning a coevolution scheme between host galaxies and AGNs \citep[e.g.,][]{2013ARA&A..51..511K}, such as M$_\mathrm{BH}$--$\sigma_\star$ \citep[e.g.,][]{2000ApJ...539L...9F} and AGN feedback \citep[e.g.,][]{2006MNRAS.370..645B}. 
In addition, some studies have indicated that CT-AGNs tend to be hosted in galaxies with a significant amount of dense gas \citep[e.g.,][]{2015ApJ...814..104K,2017MNRAS.468.1273R}. The radiation from AGNs can compress the surrounding gas, potentially triggering star formation activity within their host galaxies. \cite{2006ApJS..163....1H} suggested that the host galaxies of CT-AGNs appear to be in intense star formation. \cite{2012ApJ...755....5G} demonstrated that the host galaxies of CT-AGNs in the local Universe exhibit a high level of star formation. However, \cite{2021ApJ...908..169G} indicate similar distributions in terms of stellar masses and SFRs between the host galaxies of CT-AGNs and non-CT-AGNs in their sample. Despite the findings, the sample size of CT-AGNs is too small to ensure that the conclusions drawn are statistically robust. Similarly, we compared the SFRs and stellar masses of host galaxies for both CT-AGNs and non-CT-AGNs in our sample.

Figure~\ref{fig:SFR-M} presents relationships, known as the main sequence, between SFRs and stellar masses of the host galaxies for both CT-AGNs and non-CT-AGNs.
We also used the Kolmogorov-Smirnov test (KS test) to examine the distributions of stellar masses for their host galaxies, and the $p$ value was 0.47. We repeated the KS test to examine their SFR distributions, and the $p$ value was 0.53. 
These imply that there is no significant discrepancy in the distribution of stellar masses and SFRs among the host galaxies of different AGNs. Our findings align with the claim of \cite{2021ApJ...908..169G}, yet they contradict the conclusions drawn by \cite{2012ApJ...755....5G}. 
A potential explanation is that the environments of the AGN host galaxies at high redshifts may be distinct from those in the local Universe because our sample and that of \cite{2021ApJ...908..169G} consist of sources at high redshifts. 
A recent work by \cite{2022A&A...666A..17G} suggests that the host galaxies of high-redshift AGNs are indeed different from those of AGNs in the local Universe, as they tend to be hosted in compact galaxies with a denser interstellar medium.
Moreover, the biases in identifying CT-AGNs could potentially impact our results.
In subsequent work, we shall find more CT-AGNs using other methods in COSMOS and further study whether there are differences in the host galaxies of both CT-AGNs and non-CT-AGNs.

\section{Summary}\label{sec:summary}

In this work, we have used multiwavelength approaches to identify CT-AGNs with low photon counts in COSMOS. Firstly, we constructed a sample containing 440 AGNs with abundant multiwavelength data. Then, we analyzed multiwavelength data to extract the essential physical parameters necessary for the subsequent diagnosis of CT-AGNs, such as 6~\micron~luminosities of AGNs and the luminosities of high-ionization emission lines. 
After that, through MIR diagnosis, three sources could be identified as CT-AGNs. With the high-ionization emission lines as a diagnostic tool, 16 sources were diagnosed as CT-AGNs. In total, 18 CT-AGNs with low photon counts were identified. 
Among them, only a CT-AGN (XID \textit{lid\_673}) could be identified by both methods.
Moreover, four AGNs exhibit discrepant results when diagnosed by the two methods. Subsequently, we sought to explore the potential reasons behind the diagnostic discrepancies in the four AGNs. We further discussed the impact of estimating [O~III]$\lambda~5007$ luminosities based on [O~II]$\lambda~3727$ luminosities for the diagnosis of CT-AGNs and found that some CT-AGNs might be missed. We also saw that our process for identifying CT-AGNs contains a notable bias. This suggests that many potential CT-AGNs with low photon counts have not yet been identified.
Finally, we compared the properties of host galaxies for CT-AGNs and non-CT-AGNs in our sample and did not find significant discrepancies in the properties of their host galaxies.

\section*{Data availability}
Table~\ref{Tab:summary} is only available in electronic form at the CDS via anonymous ftp to cdsarc.u-strasbg.fr (130.79.128.5) or via http://cdsweb.u-strasbg.fr/cgi-bin/qcat?J/A+A/.

\begin{acknowledgements}
          We sincerely thank the anonymous referee for useful suggestions.
      We acknowledge the support of \emph{National Natural Science Foundation of China} (Nos  12303017).
      This work is also supported by \emph{Anhui Provincial Natural Science Foundation} project number 2308085QA33.
      QSGU is supported by the National Natural Science Foundation of China (grant numbers 12192222, 12192220, and 12121003).
      Yongyun Chen is grateful for funding for the training Program for talents in Xingdian, Yunnan Province (2081450001).
      X.L.Y. acknowledges the grant from the National Natural Science Foundation of China (NSFC grants 12303012), Yunnan Fundamental Research Projects (No.202301AT070242).
      H.T.W is supported by the Hebei Natural Science Foundation of China (Grant No. A2022408002).
\end{acknowledgements}

\bibliographystyle{aa} 
\bibliography{bibtex.bib} 

\begin{appendix} 
        
\section{Physical properties of our AGN sample}

        \begin{sidewaystable*}
                \caption{Physical properties of our AGN sample.}\label{Tab:summary}
                \centering
                \resizebox{\hsize}{!}{
                        \begin{tabular}{lcccccccccccccccc}
                                \hline\hline             
                                XID & R.A. & Decl. & z & z\_type&z\_ref&$\log$ L$_{X,2-10\mathrm{keV}}$&$\log \nu \mathrm{L}_{\nu}$(6~\micron)&$\alpha_\mathrm{AGN}$&$\log \mathrm{M_\star}$&$\log \mathrm{SFR}$&$\log \mathrm{L}_{[\mathrm{O~III}]}$&S/N$_{[\mathrm{O~III}]}$&$\log \mathrm{L}_{[\mathrm{O~II}]}$&S/N$_{[\mathrm{O~II}]}$& $\log \mathrm{L}_{[\mathrm{Ne~V}]}$ &S/N$_{[\mathrm{Ne~V}]}$\\
                                &($^\circ$)&($^\circ$)&&&&($\mathrm{erg\ s^{-1}}$)&($\mathrm{erg\ s^{-1}}$)&&($\mathrm{M}_\odot$) &($\mathrm{M}_\odot \ \mathrm{yr}^{-1}$) & ($\mathrm{erg\ s^{-1}}$)&&($\mathrm{erg\ s^{-1}}$) &&($\mathrm{erg\ s^{-1}}$)& \\
                                (1)&(2)&(3)&(4)&(5)&(6)&(7)&(8)&(9)&(10)&(11)&(12)&(13)&(14)&(15)&(16)&(17)\\
                                \hline
                                cid\_1005 & 149.6727 & 2.0058 & 2.227 & photo & M16 & -43.967 & 44.361 & 0.00002 & 10.447 & 1.685 & $\cdots$ & $\cdots$ & $\cdots$ & $\cdots$ & $\cdots$ & $\cdots$ \\
                                cid\_1007 & 149.6900 & 2.0391 & 1.809 & photo & M16 & 43.554 & 43.154 & 0.00008 & 10.833 & 0.875 & $\cdots$ & $\cdots$ & $\cdots$ & $\cdots$ & $\cdots$ & $\cdots$ \\
                                cid\_1051 & 150.2529 & 1.8711 & 1.376 & spec & H18 & 43.476 & 44.476 & 1.00000 & 10.988 & 1.727 & $\cdots$ & $\cdots$ & 42.590 & 4.72 & $\cdots$ & $\cdots$ \\
                                cid\_1074 & 149.8269 & 2.0150 & 0.750 & my-photo & This-work & -42.731 & 43.054 & 0.95907 & 9.841 & 0.794 & $\cdots$ & $\cdots$ & $\cdots$ & $\cdots$ & $\cdots$ & $\cdots$ \\
                                cid\_1168 & 150.2037 & 1.9833 & 0.600 & my-photo & This-work & -42.860 & 41.543 & 1.00000 & 7.971 & -1.484 & 41.173 & 0.39 & $\cdots$ & $\cdots$ & $\cdots$ & $\cdots$ \\
                                cid\_1305 & 150.0450 & 2.2178 & 0.828 & spec & H18 & -43.206 & 43.302 & 0.82546 & 9.669 & 1.421 & 41.281 & 0.19 & 41.292 & 0.57 & $\cdots$ & $\cdots$ \\
                                cid\_1467 & 149.8375 & 1.9718 & 1.020 & spec & H18 & 42.990 & 43.822 & 0.02199 & 10.762 & 1.948 & 43.365 & 1.01 & 44.307 & 2.54 & $\cdots$ & $\cdots$ \\
                                cid\_1474 & 149.9241 & 1.8893 & 1.551 & spec & H18 & 43.392 & 44.429 & 0.99999 & 10.369 & 1.669 & $\cdots$ & $\cdots$ & 44.810 & 2.25 & 44.667 & 0.99 \\
                                cid\_1482 & 150.1530 & 1.8805 & 1.622 & spec & H18 & -43.981 & 43.877 & 0.00006 & 10.529 & 1.556 & $\cdots$ & $\cdots$ & 42.977 & 4.81 & 42.389 & 0.43 \\
                                cid\_2728 & 150.2385 & 2.0562 & 1.506 & spec & H18 & 43.613 & 44.825 & 1.00000 & 11.069 & 1.961 & $\cdots$ & $\cdots$ & 44.348 & 6.85 & 44.292 & 5.43 \\
                                cid\_3122 & 149.9663 & 2.0328 & 0.728 & spec & H18 & -43.218 & 43.437 & 1.00000 & 11.076 & 0.899 & 41.790 & 1.21 & 42.093 & 3.38 & $\cdots$ & $\cdots$ \\
                                cid\_3242 & 149.7112 & 2.1452 & 1.531 & spec & H18 & -43.809 & 44.956 & 1.00000 & 11.231 & 2.472 & $\cdots$ & $\cdots$ & 43.212 & 1.82 & 43.785 & 0.92 \\
                                cid\_3371 & 149.9374 & 1.7390 & 0.864 & spec & H18 & 43.065 & 42.444 & 0.00003 & 10.708 & 0.992 & 43.041 & 0.16 & 43.978 & 2.24 & $\cdots$ & $\cdots$ \\
                                cid\_3570 & 149.6409 & 2.1077 & 1.244 & spec & H18 & 43.291 & 44.276 & 1.00000 & 11.006 & 1.475 & $\cdots$ & $\cdots$ & 43.124 & 14.07 & $\cdots$ & $\cdots$ \\
                                cid\_3714 & 150.2400 & 2.3533 & 1.265 & spec & H18 & 43.284 & 43.720 & 0.00000 & 11.069 & 2.179 & $\cdots$ & $\cdots$ & 43.288 & 6.94 & $\cdots$ & $\cdots$ \\
                                \multicolumn{1}{c}{$\vdots$}&$\vdots$&$\vdots$&$\vdots$&$\vdots$&$\vdots$&$\vdots$&$\vdots$&$\vdots$&$\vdots$&$\vdots$&$\vdots$&$\vdots$&$\vdots$&$\vdots$&$\vdots$&$\vdots$ \\
                                lid\_611 & 149.8474 & 2.5692 & 0.913 & spec & H18 & 42.808 & $\cdots$ & 0.00000 & 11.070 & 2.037 & $\cdots$ & $\cdots$ & 43.666 & 1.38 & $\cdots$ & $\cdots$ \\
                                lid\_617 & 149.6254 & 2.3492 & 1.640 & spec & H18 & 43.257 & 43.660 & 0.00000 & 11.178 & 1.812 & $\cdots$ & $\cdots$ & 44.371 & 0.77 & $\cdots$ & $\cdots$ \\
                                lid\_673 & 149.6210 & 2.0795 & 1.452 & spec & H18 & 43.293 & 45.126 & 1.00000 & 10.953 & 1.086 & $\cdots$ & $\cdots$ & 42.650 & 5.00 & $\cdots$ & $\cdots$ \\
                                lid\_943 & 149.5330 & 1.9588 & 0.565 & spec & H18 & 42.338 & 43.577 & 1.00000 & 9.835 & -0.979 & $\cdots$ & $\cdots$ & 42.827 & 1.31 & $\cdots$ & $\cdots$ \\
                                lid\_950 & 149.5122 & 1.9654 & 2.446 & photo & M16 & 43.648 & 45.109 & 1.00000 & 11.238 & 2.257 & $\cdots$ & $\cdots$ & $\cdots$ & $\cdots$ & $\cdots$ & $\cdots$ \\
                                lid\_991 & 149.7587 & 1.7552 & 0.831 & photo & M16 & 43.002 & 43.359 & 1.00000 & 10.386 & 0.036 & $\cdots$ & $\cdots$ & $\cdots$ & $\cdots$ & $\cdots$ & $\cdots$ \\
                                
                                \hline
                        \end{tabular}
                }
                \tablefoot{
                	\textbf{This table is available in its entirety at the CDS.} Col. (1) contains X-ray IDs from the \cite{2016ApJ...819...62C} catalog. Cols. (2) and (3) contain the R.A. and decl. of the X-ray sources from the \cite{2016ApJ...819...62C} catalog. Col. (4) contains the redshift that we used. Col. (5) contains the type of redshift, including "spec", "photo" and "my-photo". The "spec" represent spectral redshift. The "photo" represent photometric redshift from \citetalias{2016ApJ...817...34M}.  The "my-photo" is photometric redshift which obtained by SED fitting in this work. Col. (6) lists the references that provide this redshift. "H18" represent \cite{2018ApJ...858...77H}. Col. (7) contains the logarithm of observed 2-10 keV luminosity in the rest-frame. Among them, negative value represents the upper limit of its luminosity. Col. (8) contains the logarithm of 6~\micron~luminosity for an AGN in the rest-frame. Col. (9) contains the confidence of AGN components for SED fitting. Cols. (10) and (11) contain the logarithm of M$_\star$ and SFR for AGN host galaxy. Cols. (12), (14) and (16) contain the logarithm of luminosities for [O~III], [O~II] and [Ne~V] emission lines, respectively. Cols. (13), (15) and (17) contain the S/Ns for [O~III], [O~II] and [Ne~V] emission lines, respectively.}
        \end{sidewaystable*}

        \section{Two examples of using our photometric redshifts}\label{sec:my-photoz}
\begin{figure*}
                \includegraphics[width=0.49\linewidth]{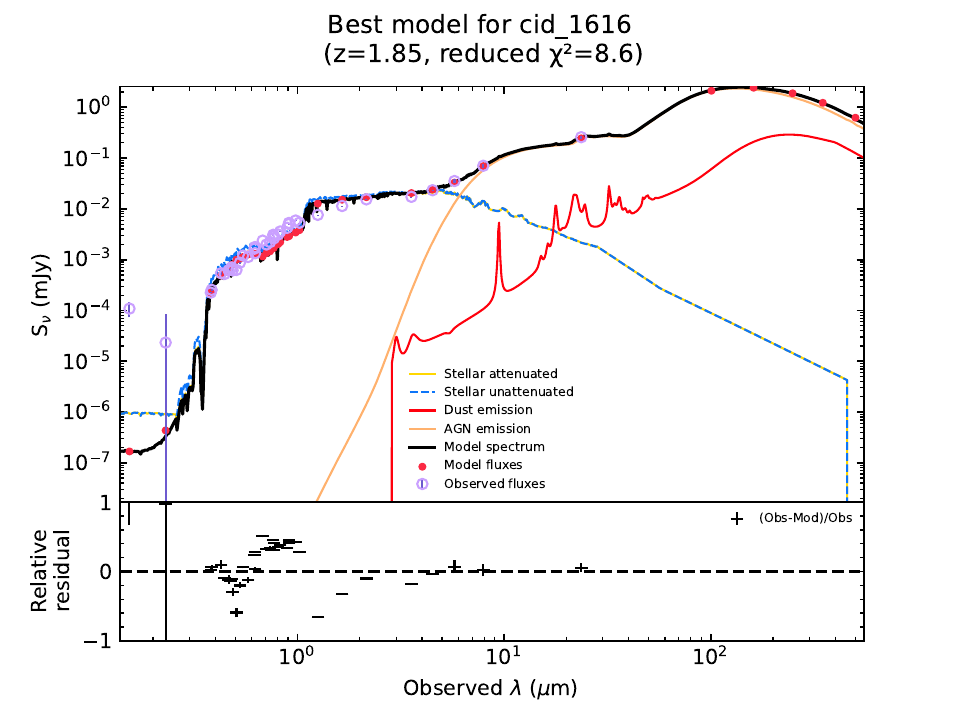}
                \includegraphics[width=0.49\linewidth]{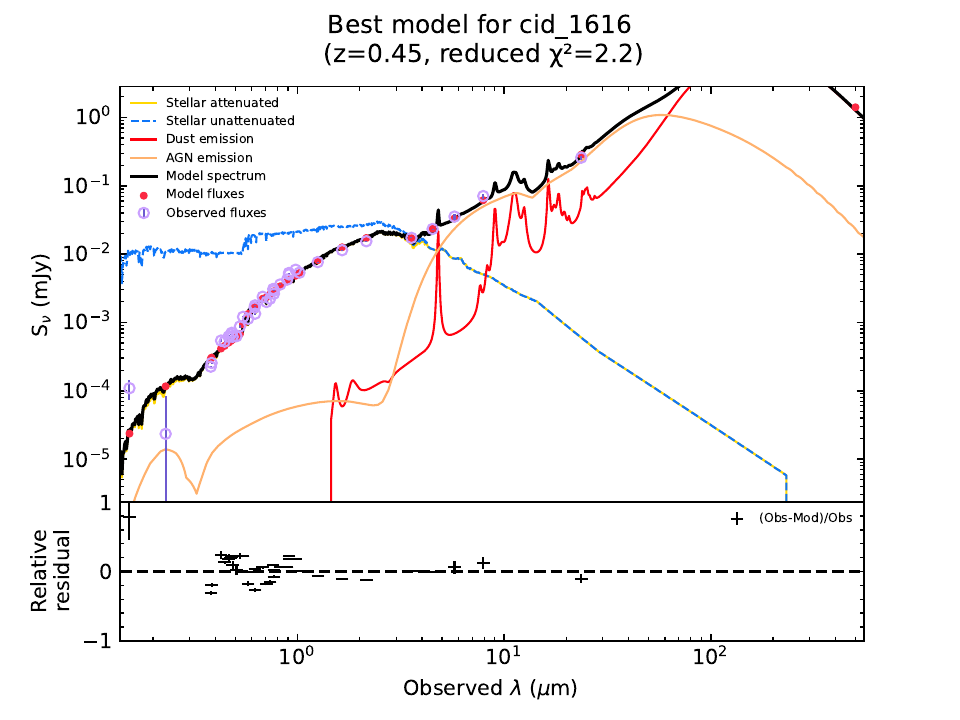}
                \includegraphics[width=0.49\linewidth]{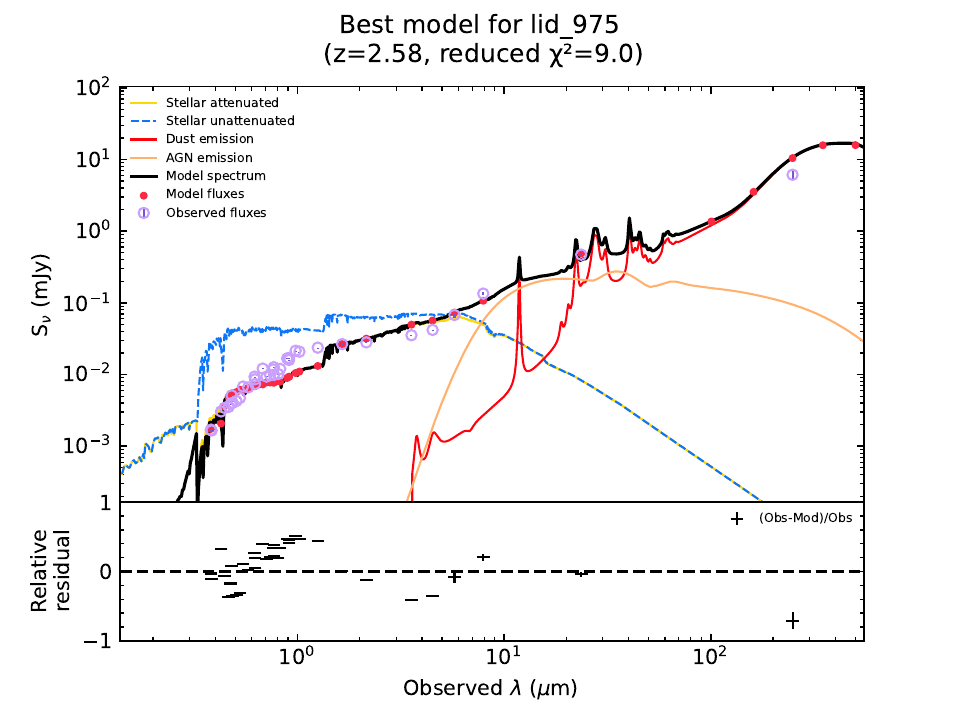}
                \includegraphics[width=0.49\linewidth]{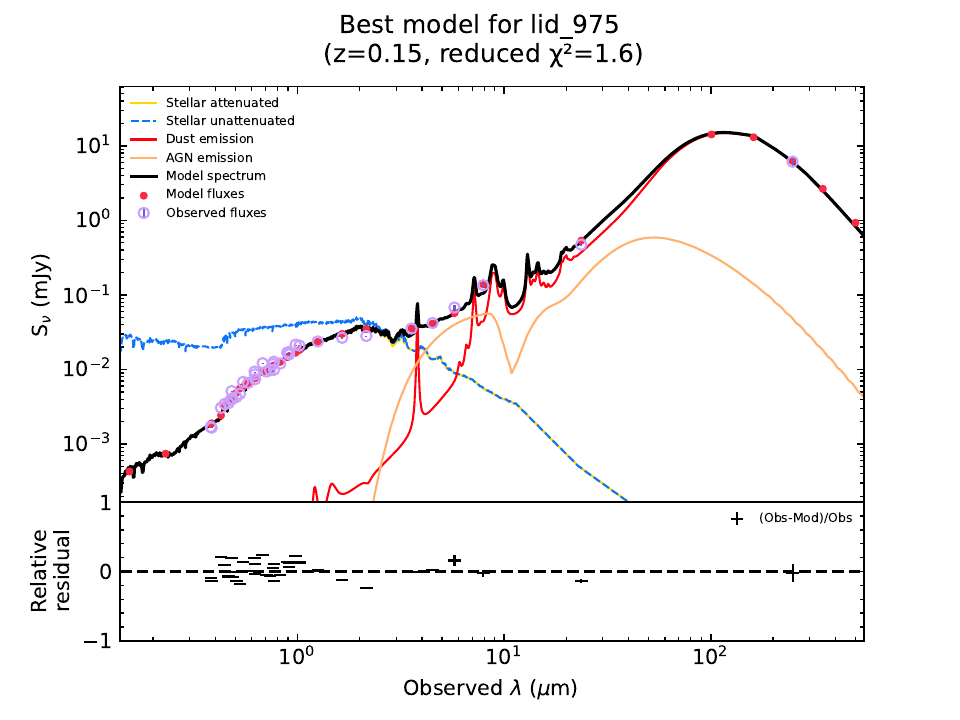}
                \caption{ SEDs fit using the photometric redshifts from the \citetalias{2016ApJ...817...34M} catalog (left) and using our photometric redshifts (red).} 
                \label{fig:my-photoz}
        \end{figure*}

        We provide examples with the XIDs \textit{cid\_1616} and \textit{lid\_975}. Their redshifts in the M16 catalog are 1.852 and 2.581, respectively, while our photometric redshifts are 0.45 and 0.15. The left panels of Fig.~\ref{fig:my-photoz} show the SEDs fit using the redshifts from the \citetalias{2016ApJ...817...34M} catalog, and the right panels display the SEDs fit with our redshifts.
        
        \section{Confidence coefficient of AGN components}\label{sec:quantifying}
        BIC is a statistical method for selecting the best model from a limited set. The values of BIC for a model are given by
        \begin{equation}
                \label{equ:BIC}
                BIC = -2\ln(L) + p\ln(N),
        \end{equation}
        where $L$ is the maximum likelihood of the model, $p$ and $N$ are the number of parameters and the number of observations, respectively. When fitting data with errors, assuming the errors are independent and normally distributed, the likelihood function can be written as
        \begin{equation}
                \label{equ:likelihood}
                \begin{split}
                        L =& \prod_{i=1}^N \frac{1}{\sqrt{2\pi \sigma_i}} \exp\left(-\frac{\left(y_i-f(x_i) \right)^2}{2\sigma_i^2}\right)  \\ 
                        =& \left( \frac{1}{\sqrt{2\pi}} \right)^N  \left(\prod_{i=1}^N \sigma_i \right)  \exp\left(-\frac{1}{2} \sum_{i=1}^{N} \frac{\left(y_i-f(x_i) \right)^2}{\sigma_i^2}\right),
                \end{split}
        \end{equation}
        where $y_i$ are the observed data points, $x_i$ are the independent variables, $f(x)$ is the model function, and $\sigma_i$ are the standard deviations of the errors. Among them, 
        \begin{equation}
                \label{equ:chi-square}
                \sum_{i=1}^{N} \frac{\left(y_i-f(x_i) \right)^2}{\sigma_i^2} = \chi^2.
        \end{equation}
        Substituting Eqs.~\eqref{equ:likelihood} and~\eqref{equ:chi-square} into Eq.~\eqref{equ:BIC} gives:
        \begin{equation*}
                BIC = N \ln(2\pi) + \sum_{i=1}^{N} \sigma_i + \chi^2 + p\ln(N).
        \end{equation*}
        
        To compare two models (with or without \textit{AGN} model) and select the best one, we calculate the information:
        \begin{equation*}
                \Delta BIC = \chi_2^2 -\chi_1^2+ (p_2-p_1)\ln(N),
        \end{equation*}
        where $\chi_2^2$ and $p_2$ are the parameters with \textit{galaxy+AGN} model, $\chi_1^2$ and $p_1$ are the parameters with \textit{galaxy} model. We further consider the Schwarz weights \citep{2002Model}:
        \begin{equation}
                \mathrm{weight} = \frac{\exp\left(-\frac{1}{2} \Delta BIC \right)}{\exp\left(\frac{1}{2} \Delta BIC \right)+\exp\left(-\frac{1}{2} \Delta BIC \right)}.
        \end{equation}
        These weights are a quantitative measure to assess the probability of including AGN components in SED fitting. In other words, they express a confidence coefficient of AGN components obtained through SED fitting, that is $\alpha_\mathrm{AGN} = \mathrm{weight}$.
        \section{X-ray absorbtion for CT-AGNs}\label{sec:absorb}
        The X-ray radiation from AGNs is complex. It typically includes several distinct components within their spectra: initial X-ray emission from the corona, reflection component from the accretion disk, soft excess component, scattering from the polar regions, and diffuse thermal emission. For CT-AGNs, the initially soft X-ray radiation is almost entirely absorbed by the dust torus. The polar scattering components, however, are not obscured by the torus. Consequently, when considering the radiation of CT-AGNs, it must take into account the contribution of the polar scattering component. To determine the (flux or luminosity) ratio of X-ray radiation from AGNs both before and after significant Compton-thick absorption, we consider three components: the initial X-ray emission, reflection from the accretion disk, and polar scattering. So, the model in the \textit{Xspec} terminology is written as follows: 
        \begin{align*}
                \mathrm{model =}& \mathrm{phabs[1]*powerlaw[2] + constant[3]*powerlaw[4]} \\
                & \mathrm{+ pexmon[5]}
        \end{align*}
        The detailed parameters of this model are listed in Table~\ref{table:absorbtion}. Based on this model, the flux or luminosity ratio of 2--10~keV before and after significant Compton-thick absorption is 17.625.
        
        \begin{table}
                \caption{Parameters of the X-ray model.} \label{table:absorbtion}
                \centering
                \resizebox{\hsize}{!}
                {
                        \begin{tabular}{l c c c c}
                                \hline\hline
                                Component & Parameter & Unit & \multicolumn{2}{c}{Value} \\ 
                                \cline{4-5}
                                &&&Unabsorbed&Compton-thick\\
                                (1)&(2)&(3)&(4)&(5)\\
                                \hline
                                phabs[1] & N$_\mathrm{H}$ & $10^{22} \ \mathrm{cm}^{-2}$ & 0 &150 \\
                                \cline{1-5}
                                \multirow{2}{*}{powerlaw[2]}& PhoIndex & &1.9&1.9\\
                                & norm & & 1& 1\\
                                \cline{1-5}
                                constant[3] & factor & & 0.005 &0.005 \\
                                \cline{1-5}
                                \multirow{2}{*}{powerlaw[4]}& PhoIndex & &1.9&1.9\\
                                & norm & & 1& 1\\
                                \cline{1-5}
                                \multirow{5}{*}{pexmon[5]}& PhoIndex & &1.9&1.9\\
                                & foldE & keV & 500& 500\\
                                & redshift & & 0& 0\\
                                & Incl & deg& 85& 85\\
                                & norm & & 1& 1\\
                                \hline
                        \end{tabular}
                }
        \end{table}
        
\end{appendix}

\end{document}